\documentclass[prl,twocolumn,floats,aps,showpacs,superscriptaddress,nofootinbib]{revtex4}

\usepackage[dvips]{graphicx}
\usepackage{dcolumn}
\usepackage{epsfig}

\begin{document}
\title{Disentangling the dynamical origin of $P_{11}$ Nucleon Resonances}
\author{N. Suzuki}
\affiliation{Department of Physics, Osaka University, Toyonaka,
Osaka 560-0043, Japan}
\affiliation{Excited Baryon Analysis Center (EBAC), Thomas Jefferson National
Accelerator Facility, Newport News, VA 23606, USA}
\author{B. Juli\'a-D\'{\i}az}
\affiliation{Department d'Estructura i Constituents de la Mat\`{e}ria
and Institut de Ci\`{e}ncies del Cosmos,
Universitat de Barcelona, E--08028 Barcelona, Spain}
\affiliation{Excited Baryon Analysis Center (EBAC), Thomas Jefferson National
Accelerator Facility, Newport News, VA 23606, USA}
\author{H. Kamano}
\affiliation{Excited Baryon Analysis Center (EBAC), Thomas Jefferson National
Accelerator Facility, Newport News, VA 23606, USA}
\author{T.-S. H. Lee}
\affiliation{Excited Baryon Analysis Center (EBAC), Thomas Jefferson National
Accelerator Facility, Newport News, VA 23606, USA}
\affiliation{Physics Division, Argonne National Laboratory,
Argonne, IL 60439, USA}
\author{A. Matsuyama}
\affiliation{Department of Physics, Shizuoka University, Shizuoka 422-8529, Japan}
\affiliation{Excited Baryon Analysis Center (EBAC), Thomas Jefferson National
Accelerator Facility, Newport News, VA 23606, USA}
\author{T. Sato}
\affiliation{Department of Physics, Osaka University, Toyonaka,
Osaka 560-0043, Japan}
\affiliation{Excited Baryon Analysis Center (EBAC), Thomas Jefferson National
Accelerator Facility, Newport News, VA 23606, USA}

\begin{abstract}
We show that two almost degenerate poles near the $\pi\Delta$ threshold
and the next higher mass pole in the $P_{11}$ partial wave of $\pi N$ 
scattering evolve from a single bare state through its coupling with 
$\pi N$, $\eta N$ and $\pi\pi N$ reaction channels. This finding provides 
new information on understanding the dynamical origins of the Roper 
$N^*(1440)$ and $N^*(1710)$ resonances listed by Particle Data Group.
Our results for the resonance poles in other $\pi N$ partial waves 
are also presented. 
\end{abstract}
\pacs{14.20.Gk, 13.75.Gx, 13.60.Le}

\maketitle

The excited nucleon states are unstable and couple strongly to 
the meson-baryon continuum states to form resonances in $\pi N$ 
and $\gamma N$ reactions. Therefore, the extraction of nucleon 
resonances (called collectively as $N^*$) from data has been a 
well recognized important task in advancing our understanding of 
strong interactions. The $N^*$ parameters listed and periodically 
updated by Particle Data Group~\cite{pdg} (PDG) are commonly 
used in testing hadron structure calculations using QCD-based 
hadron models~\cite{isgur79,riska,roberts} and Lattice 
QCD~\cite{lqcd,lattice2}.

It is well known that resonances locate on the unphysical
sheets of the complex energy plane and thus their properties 
can only be extracted from the empirical partial-wave amplitudes 
(PWA) by analytic continuation. In extracting resonances from 
$\pi N$ data up to invariant mass W=2 GeV we face a multi-channel 
complication, namely that a resonance may appear as a pole on 
more than one of the unphysical Riemann sheets, as investigated 
previously by Eden and Taylor~\cite{et63}, Kato~\cite{kato65}, 
and Morgan and Pennington~\cite{mp87}. It is custom to name the 
pole which is closest to physical region as the resonance pole, 
and others as shadow poles. In general, the observables are 
mainly determined by the resonance poles. However, under certain 
circumstances a shadow pole could lie close to the threshold of 
one of the channels and could therefore affect the physical 
observables, as discussed in Refs. ~\cite{et63,mp87}. A theoretical 
understanding of the dynamical origins of these poles and their 
inter-relations is needed to interpret the resonance parameters. 
In this letter, we report a progress in this direction for the
$N^*$ in the $P_{11}$ partial wave of $\pi N$ scattering.
Our results for other partial waves will also be presented.

The determination of resonance poles in the $P_{11}$ partial wave 
has been difficult since the discovery~\cite{roper64} of the 
Roper, $N^*$(1440), resonance in 1964. It was first found by Arndt, Ford 
and Roper~\cite{afr84} that this partial wave has two almost 
degenerate poles near the $\pi \Delta$ threshold. This was confirmed 
and investigated in more detail by Cutkosky and Wang~\cite{cw90}.
This two pole structure has also been obtained in the recent 
analysis by the GWU/VPI~\cite{said-1} and J\"{u}lich~\cite{juelich} 
groups. In this letter, we demonstrate that these two poles near 
the $\pi\Delta$ threshold ($\sim $ 1360 MeV) and a pole at about 1800 
MeV correspond to a single bare state within a dynamical coupled-channels 
model (JLMS) developed in Ref.~\cite{jlms07}. Thus they have the 
resonance pole-shadow pole relation as discussed in 
Refs.~\cite{et63,kato65,mp87}. Our result suggest that the 
$N^*$(1440) and $N^*(1710)$ listed by PDG originate from the same 
excited nucleon state modeled as a bare particle within the JLMS model.

The JLMS model is defined within a Hamiltonian formulation~\cite{msl} of 
multi-channels reactions. It describes meson-baryon ($MB$) reactions 
involving the following channels: $\pi N$, $\eta N$, and $\pi\pi N$ which 
has $\pi\Delta$, $\rho N$, and $\sigma N$ resonant components. The 
excitation of the internal structure of a baryon ($B$) by a meson 
($M$) to a bare $N^*$ state is modeled by a vertex interaction 
$\Gamma_{MB \leftrightarrow N^*}$. The Hamiltonian also has energy 
independent interactions $v_{MB,M'B'}$ which describe the meson-exchange 
mechanisms deduced from phenomenological Lagrangians. Nucleon resonances 
can be due to the $MB \to N^* \to M'B'$ transitions induced by the 
vertex interaction $\Gamma_{MB \leftrightarrow N^*}$ in this formulation.
But they can also be due to the attractive forces of  $v_{MB,M'B'}$ and
channel coupling effects. For investigating the $N^*$ structure, the
second type of resonances, called molecular-type resonances in the literature,
must also be identified in the analysis.
For the same consideration, the parameterization of $v_{MB,M'B'}$,
in particular their phenomenological form factors, must be carefully
constrained by the data. This had been achieved
 by performing rather complex $\chi^2$-fits 
to the $\pi N$ scattering data, as detailed
in Ref.~\cite{jlms07}. Briefly, the JLMS model is able to describe the data of
$\pi N$ elastic scattering up to invariant mass $W=$ 2 GeV.
The resulting $\pi N$ scattering amplitudes and total cross sections
are in good agreement with
those from SAID~\cite{said-1}.
 Furthermore, the predicted $2\pi$ production cross sections~\cite{kjlms09}
are in good agreement with the available data. 

Within the JLMS model, it is convenient to cast the partial-wave amplitude 
of the $M(\vec k)+ B(-\vec k) \to M'(\vec k')+ B'(-\vec k')$ reaction
into the following form (suppressing the angular momentum and isospin
indices):
\begin{eqnarray}
T_{MB,M'B'}(k,k',E)  &=&  t_{MB,M'B'}(k,k',E) \nonumber\\
&+& t^{N^*}_{MB,M'B'}(k,k',E) \,,
\label{eq:tmbmb}
\end{eqnarray}
where the first term (called meson-exchange amplitude from now on) is defined by
\begin{eqnarray}
  t_{MB,M^\prime B^\prime}(k,k',E)&=&  v_{MB,M^\prime B^\prime}(k,k')
\label{eq:cc-mbmb}
\\
&+& \sum_{M^{\prime\prime}B^{\prime\prime}}
\int_{C_{M^{\prime\prime}B^{\prime\prime}}} \!\!\!\!\!\!\!\!\!\! q^2\; dq\;
v_{MB,M^{\prime\prime}B^{\prime\prime}}(k,q)
 \nonumber \\
 &\times&
G_{M^{\prime\prime}B^{\prime\prime}}(q,E)
t_{M^{\prime\prime}B^{\prime\prime},M^\prime B^\prime}(q,k'E)\,, 
\nonumber 
\end{eqnarray}
where $C_{MB}$ is the integration contour in the complex$-q$ plane used 
for channel $MB$.
The term associated with the bare $N^*$ states in Eq.~(\ref{eq:tmbmb}) is
\begin{eqnarray}
t^{N^*}_{MB,M^\prime B^\prime}(k,k',E)&=& \sum_{N^*_i, N^*_j}
\bar{\Gamma}_{MB \to N^*_i}(k,E) [D(E)]_{i,j}
\nonumber \\
&\times&
\bar{\Gamma}_{N^*_j \to M^\prime B^\prime}(k',E),
\label{eq:tmbmb-r}
\end{eqnarray}
where $\bar{\Gamma}_{N^*_j \to M^\prime B^\prime}(k,E)$ is the
dressed vertex function which is calculated~\cite{jlms07} from 
the bare vertex ${\Gamma}_{N^*_j \to M^\prime B^\prime}(k)$ and convolutions
over the meson-exchange amplitudes $t_{MB,M^\prime B^\prime}(k,k',E)$.
The inverse of the propagator of dressed $N^*$ states in
Eq.~(\ref{eq:tmbmb-r}) 
is \begin{eqnarray}
[D^{-1}(E)]_{i,j} &=& (E - M^0_{N^*_i})\delta_{i,j} - [M(E)]_{i,j} \, ,
\label{eq:nstar-selfe}
\end{eqnarray}
where $M^0_{N^*_i}$  is the bare mass of the $i$-th $N^*$ state, 
and
the $N^*$ self-energy is defined by
\begin{eqnarray}
[M(E)]_{i,j}&=&
\sum_{MB}
\int_{C_{MB}} \!\!\!\! q^2 dq 
\bar{\Gamma}_{N^*_j \to M B}(q,E)
\nonumber \\ 
&\times&
G_{MB}(q,E)\,
{\Gamma}_{MB \to N^*_i}(q,E)\,.  
\label{eq:nstar-g}
\end{eqnarray}
Defining  $E_\alpha(k)=[m^2_\alpha + k^2]^{1/2}$ with $m_\alpha$ being
the mass of particle $\alpha$, 
the meson-baryon propagators in the above equations are:
$G_{MB}(k,E)=1/[E-E_M(k)-E_B(k) + i\epsilon]$ for the stable 
$\pi N$ and $\eta N$ channels, and $G_{MB}(k,E)=1/[E-E_M(k)-E_B(k) -\Sigma_{MB}(k,E)]$
for the unstable $\pi\Delta$, $\rho N$, and $\sigma N$ channels. 
The self energy $\Sigma_{MB}(k,E)$ is calculated from a vertex 
function defining the decay of the considered unstable particle 
in the presence of a spectator $\pi$ or $N$ with momentum $k$. 
For example, we have for the $\pi\Delta$ state,
\begin{eqnarray}
\Sigma_{\pi\Delta}(k,E) &=&\frac{m_\Delta}{E_\Delta(k)}
\int_{C_3} q^2 dq \frac{ M_{\pi N}(q)}{[M^2_{\pi N}(q) + k^2]^{1/2}}
\nonumber \\
&\times&
\frac{\left|f_{\Delta \to \pi N}(q)\right|^2}
{E-E_\pi(k) -[M^2_{\pi N}(q) + k^2]^{1/2} + i\epsilon},
\label{eq:self-pid}
\end{eqnarray}
where $M_{\pi N}(q) =E_\pi(q)+E_N(q)$ and $f_{\Delta \to \pi N}(q)$
defines the decay of the $\Delta \to \pi N$ in the rest frame 
of $\Delta$, $C_3$ is the corresponding integration contour in the 
complex$-q$ plane. The self-energies for $\rho N$ and $\sigma N$ channels 
are similar.

To search for resonance poles, we need to choose the contours 
$C_{MB}$ and $C_3$ appropriately to solve Eqs.~(\ref{eq:cc-mbmb})-(\ref{eq:self-pid}) 
for $E$ on the various possible unphysical sheets of the Riemann surface. 
This requires careful examinations of the locations of the on-shell momentum  
of each propagator $G_{MB}(k,E)$ and the $\pi\pi N$ cut in the self energies, 
such as $\Sigma_{\pi\Delta}(k,E)$ of Eq.~(\ref{eq:self-pid}), of the unstable 
particle channels. Furthermore, we need to account for the singularities of 
$v_{MB,M'B'}(k,k')$ of Eq.~(\ref{eq:cc-mbmb}) on the chosen contours.  Our 
method was tested~\cite{ssl09} within several exactly solvable models. 
Like all previous works~\cite{said-1,pitt-anl}, we only look for poles 
which are close to the physical region and have effects on $\pi N$ 
scattering  observables. All of these poles are on the unphysical sheet 
of the $\pi N$ channel, but could be on either unphysical $(u)$ or 
physical $(p)$ sheets of other channels considered in this analysis.
We will indicate the sheets where the identified poles are located by      
$(s_{\pi N}, s_{\eta N}, s_{\pi \pi N} ,s_{\pi \Delta},s_{\rho N},
s_{\sigma N})$, where $s_{MB}$ and $s_{\pi\pi N}$ can be 
$u$ or $p$ or $-$ denoting no coupling to this channel. 

Eq.~(\ref{eq:tmbmb}) indicates that if no pole is found in the first term 
$t_{\pi N, \pi N}(k,k',E)$, then the poles of the total amplitude can be 
found from the second term $t^{N^\ast}_{\pi N, \pi N}(k,k',E)$. But if 
$t_{\pi N, \pi N}(k,k',E)$ has a pole, we need to check whether it will 
be canceled by the second term, as demonstrated in Ref.~\cite{juelich}. 
Thus our procedure is to first use the standard method to determine 
whether $t_{\pi N, \pi N}(k,k',E)$ has poles by examining the determinant 
of $[1-vG]^{-1}$ of Eq.~(\ref{eq:cc-mbmb}). It turns out that we don't 
find any pole from these meson-exchange amplitudes. Thus there is no
molecular-type nucleon resonance within JLMS model.

\begin{figure}[t]
\begin{center}
\includegraphics[width=7.cm]{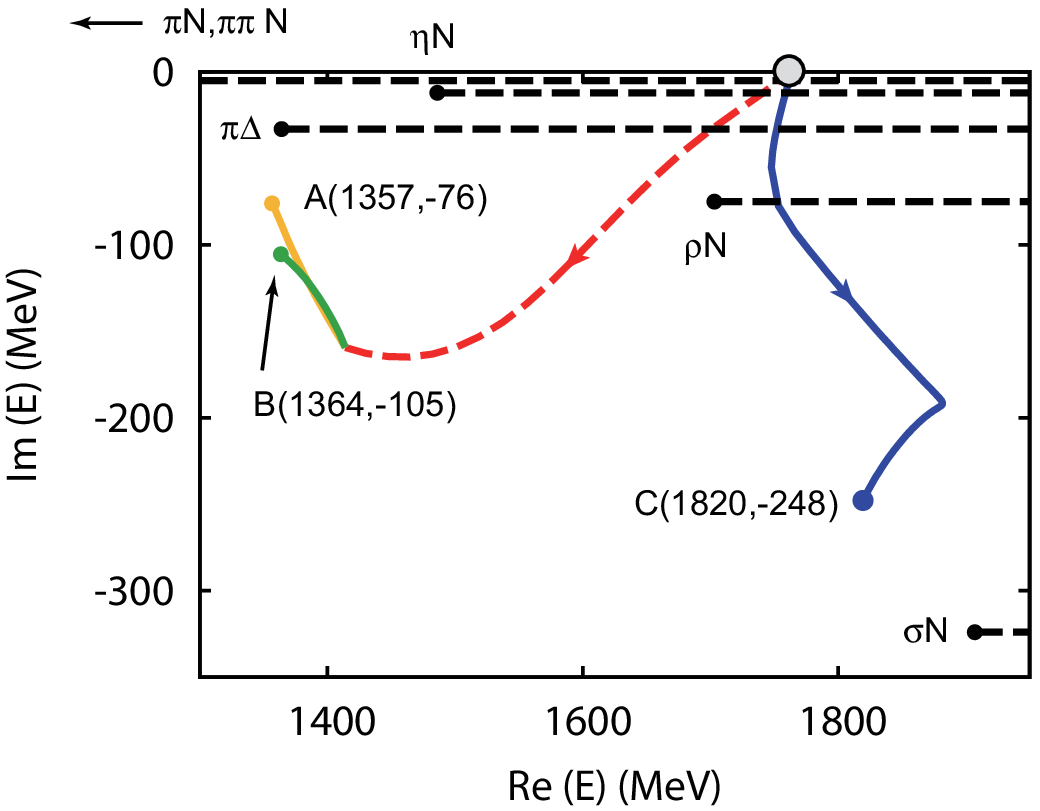}
\hspace{1cm}
\includegraphics[width=7.cm]{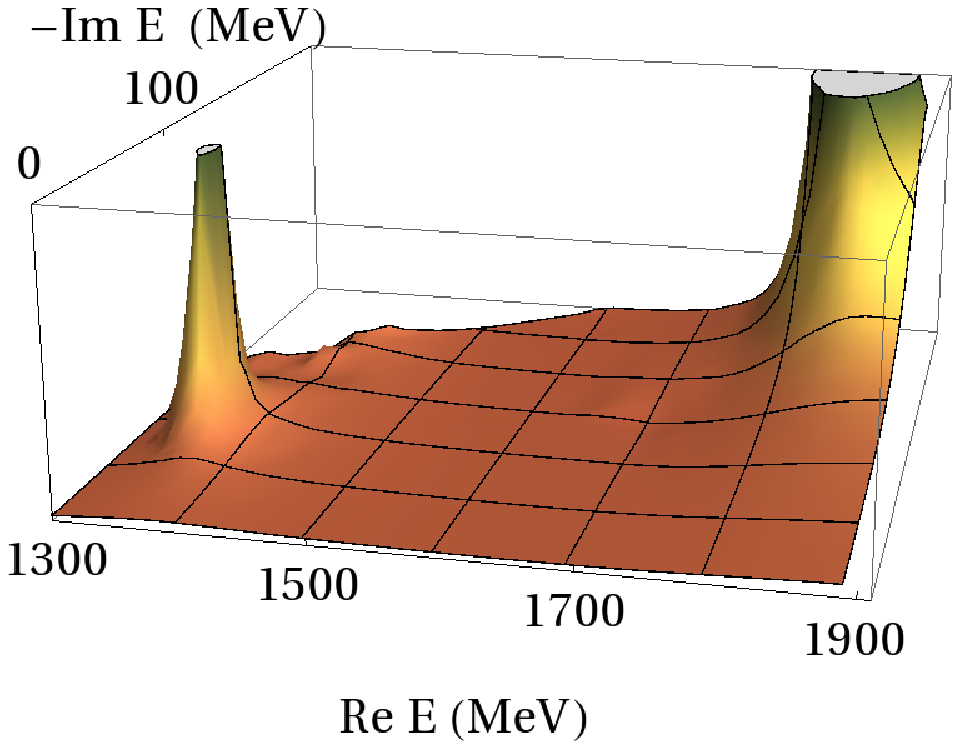}
\caption{
(above) 
Trajectories of the evolution of $P_{11}$ resonance 
poles A (1357,76), B (1364,105), and C (1820,248) from 
a bare $N^*$ with 1763 MeV, as the couplings of the 
bare $N^*$  with the meson-baryon reaction channels 
are varied from zero to the full strengths of the JLMS 
model. See text for detailed explanations. Brunch cuts for all channels
are denoted as dashed lines.
The branch points, $E_{\rm b.p.}$, for  unstable  channels are
determined by  $E_{\rm b.p.}- E_M(k)-E_B(k)- \Sigma_{MB}(k,E_{\rm b.p.})=0$ 
of the their propagators (described in the text) evaluated at the spectator 
momentum $k$=0.
With the parameters~\cite{msl} used in JLMS model, we find that
$E_{\rm b.p.}$ (MeV) $ =( 1365.40, - 32.46), (1704.08, -74.98),( 1907.57, -323.62)$
for $\pi\Delta$, $\rho N$, and $\sigma N$, respectively. 
 (below) 3-Dimensional 
depiction of the behavior of $\left|\text{det}[ D(E)]\right|^{2}$ 
of the $P_{11}$ $N^*$ propagator (in arbitrary units) as 
a function of complex-$E$.  }
\label{fig:p11-pole}
\end{center}
\end{figure}

\begin{table}[b]
\caption{$P_{11}$ resonance pole positions $M_R$ [listed as $(\text{Re}~M_R,
    -\text{Im}~M_R)$] extracted from four 
different approaches are compared.
\label{tab2}}
\begin{ruledtabular}
\begin{tabular}{ccccc}          
Analysis & \multicolumn{2}{c}{P11 poles (MeV)}\\
\hline
JLMS~\cite{jlms07}      &(1357, \, 76)&(1364,   105)\\
CMB~\cite{cw90} &(1370,\, 114)& (1360,\, 120) \\
GWU/VPI~\cite{said-1}       &(1359, \, 82)&(1388, \, 83)\\
J\"{u}lich~\cite{juelich}&(1387, \, 74)&(1387, \, 71)
\end{tabular}

\end{ruledtabular}
\end{table}

\begin{table}[t]
\caption{The resonance pole positions $M_R$ [listed as $(\text{Re}~M_R,
    -\text{Im}~M_R)$] extracted from the JLMS model in the 
different unphysical sheets are compared with the values of 3- and 4-stars
nucleon resonances listed in the PDG~\cite{pdg}. The notation indicating
their locations on the Riemann surface are explained in the text. ``---" for 
$P_{33}(1600)$, $P_{13}$ and $P_{31}$ indicates that no resonance pole has
 been found in the considered complex energy region, Re$(E)\leq 2000$ MeV 
and $-$Im$(E)\leq 250$ MeV. All masses are in MeV.
\label{tab1}}
\begin{ruledtabular}
\begin{tabular}{ccccl}      
         &$M^0_{N^*}$&$M_R$ & Location      & PDG   \\
\hline
$S_{11}$ &1800       &(1540,   191)&$(uuuupp)$ &(1490 - 1530, \, 45 -   125)\\
         &1880       &(1642, \, 41)&$(uuuupp)$ &(1640 - 1670, \, 75 - \, 90)\\
$P_{11}$ &1763       &(1357, \, 76)&$(upuupp)$ &(1350 - 1380, \, 80 -   110)\\
         &1763       &(1364,   105)&$(upuppp)$ &                            \\
         &1763       &(1820,   248)&$(uuuuup)$ &(1670 - 1770, \, 40 -   190)\\
$P_{13}$ &1711       &\multicolumn{2}{c}{---}  &(1660 - 1690, \, 57 -   138)\\
$D_{13}$ &1899       &(1521, \, 58)&$(uuuupp)$ &(1505 - 1515, \, 52 - \, 60)\\
$D_{15}$ &1898       &(1654, \, 77)&$(uuuupp)$ &(1655 - 1665, \, 62 - \, 75)\\
$F_{15}$ &2187       &(1674, \, 53)&$(uuuupp)$ &(1665 - 1680, \, 55 - \, 68)\\
$S_{31}$ &1850       &(1563, \, 95)&$(u\text{--}uup\text{--})$ &(1590 - 1610, \, 57 - \, 60)\\
$P_{31}$ &1900       &\multicolumn{2}{c}{---}  &(1830 - 1880,   100 -   250)\\
$P_{33}$ &1391       &(1211, \, 50)&$(u\text{--}ppp\text{--})$ &(1209 - 1211,
\, 49 - \, 51)\\
        &1600       & \multicolumn{2}{c}{---} &(1500 - 1700, 200 -  400)\\
$D_{33}$ &1976       &(1604,   106)&$(u\text{--}uup\text{--})$ &(1620 - 1680, \, 80 -   120) \\
$F_{35}$ &2162       &(1738,   110)&$(u\text{--}uuu\text{--})$ &(1825 - 1835,   132 -   150)\\
         &2162       &(1928,   165)&$(u\text{--}uuu\text{--})$ &                            \\
$F_{37}$ &2138       &(1858,   100)&$(u\text{--}uuu\text{--})$ &(1870 - 1890,   110 -   130)
\end{tabular}
\end{ruledtabular}
\end{table}

We thus can search for poles of the total amplitudes from finding the 
zeros of the determinant of $D^{-1}(E)$ defined by
Eq.~(\ref{eq:nstar-selfe}). 
Here we use the well-established Newton iteration method. We have 
performed searches in the $(m_\pi + m_N) \leq {\rm Re}(E) \leq 2000$ 
MeV and $-{\rm Im}(E) \leq 250 $ MeV region within which PDG's 
3- and 4-stars resonances are listed. Poles with very large widths 
are more difficult to locate precisely with our numerical methods 
and hence will not be discussed here.

We now focus on our results in $P_{11}$ partial wave. We find two 
poles near the PDG value  $(\text{Re}~M_R, -\text{Im}~M_R) =$ 
(1350$-$1380, 80$-$110) of the Roper, $N^*$(1440),
resonance. This finding is consistent with the results from 
the analysis by Cutkosky and Wang~\cite{cw90} (CMB),
GWU/VPI~\cite{said-1} and  J\"{u}lich ~\cite{juelich} groups, as 
seen in Tab.~\ref{tab2}. In our analysis, we find that they 
are on different sheets: (1357,76) and (1364,105) are on the 
un-physical and physical sheet of the $\pi\Delta$ channel, respectively.

We also find one higher mass pole at $(1820, 248)$ in $P_{11}$ partial wave,  
which is close to the $N^*(1710)$ state listed by PDG. Within 
the JLMS model, we find that this pole and the two poles listed 
in table~\ref{tab1} are related to one of the two bare states 
needed to obtain a good fit to the $P_{11}$ amplitude up to $W=2$ 
GeV, see~\cite{jlms07}. To see how these poles evolve dynamically 
through their coupling with reaction channels, we trace the 
zeros of $\text{det}[\hat{D}^{-1}(E)] = \text{det}[E- M^0_{N^*} - \sum_{MB}y_{MB}M_{MB}(E)]$ 
in the region $0 \leq y_{MB} \leq 1$, where $M_{MB}(E)$ is the contribution of
channel $MB$ to the self energy defined by Eq.~(\ref{eq:nstar-g}).
Each $y_{MB}$ is varied independently to find continuous evolution paths
through the various Riemann sheets on which our analytic continuation 
method is valid.

We find that the three poles listed in Table ~\ref{tab2} are 
associated to the bare state at 1736 MeV as shown in 
Fig.~\ref{fig:p11-pole}. The solid blue curve shows the 
evolution of this bare state to the position at C(1820, 248) on 
the unphysical sheet of the $\pi\Delta$ and $\eta N$ channels. 
The poles A(1357, 76) and B(1364,105) evolve from the same bare 
state on the physical sheet of the $\eta N$ channel. The dashed 
red curve indicates how the bare state evolves through varying 
all coupling strengths except keeping $y_{\pi\Delta}=0$, to about 
${\rm Re}(M_R)\sim 1400$ MeV. By further varying  $y_{\pi \Delta}$ to 1 of the full 
JLMS model, it then splits into two trajectories; one moves to 
pole A(1357,76) on the unphysical sheet and the other to 
B(1364, 105) on the 
physical sheet of $\pi\Delta$ channel. Fig.~\ref{fig:p11-pole} clearly 
shows how the coupled-channels effects induces multi-poles from a 
single bare state. The evolution of the second bare state at 
2037 MeV~\cite{jlms07} into a resonance at $W > 2$ GeV 
can be similarly investigated, but 
will not be discussed here.

To explore this interesting result further and to examine the stability of
the determined three $P_{11}$ poles, we have performed several refits of
the $P_{11}$  amplitudes within the JLMS model. We are able to get new fits
by varying solely the parameters associated with the bare $N^*$ state   
at 1763 MeV while keeping its bare mass value varied within the
range $1763 \pm 100 $ MeV.
The quality of these fits are comparable to that of the original JLMS 
model.
The above described features remain unchanged: we find in all refitted results
 two poles close to the $\pi \Delta$ threshold, within 1 MeV of the 
positions reported in Table II. The third higher mass pole is also found 
 but its position varies up to 30 MeV from the 
value given in Table II. The trajectories similar to that shown in 
Fig.~\ref{fig:p11-pole} are also obtained.
This is the extent to which the stability of 
the resonance pole-shadow pole relation among the three $P_{11}$ 
poles we can establish here. A more detailed analysis 
of the model dependence of our results would involve extensive refits by
varying  
the parameters associated with both the
 meson-exchange interaction $v_{MB, M'B'}$
and bare $N^*$ states in all partial waves  and can not be addressed 
here.

To further compare our $P_{11}$ poles with the $N^*(1440)$ and
$N^*(1710)$ listed by PDG, we  have applied the method explained in 
Ref.~\cite{ssl-II} to extract the residues $F=R e^{i\phi}$ which 
is related to S-matrix by $S(E) \rightarrow  1 + 2iF/(E-M_R) $ as 
$E \rightarrow M_R$.  We obtain 
$(R [\text{MeV}],\phi [\text{degrees}])= (36, -111), (64, -99), 
(20,-168)$ for the $P_{11}$ poles at 
$(\text{Re}~M_R,
    -\text{Im}~M_R) = (1357, 76)$, $(1364, 105)$,
 and $(1820, 248)$, respectively. 
The branching ratio of the $N^*$ decay into $\pi N$ channel
can then be estimated by evaluating $\eta_e\sim R/(-{\rm Im} (M_R)$.  
Our results for 
the $P_{11}$ poles at $(1357, 76)$ and $(1364, 105)$ are 
$49 \%$ and $61 \%$, respectively. These values are close to 
$60 - 70 \%$ of the $N^*(1440)$ listed by PDG. Our result
for the pole at $(1820, 248)$ is $8 \% $ which is also close to 
$ 10 -20 \%$ of $N^*(1710)$. We thus have firmer evidence showing that
these two $N^*$ states listed by PDG do evolve from the same bare state
through its coupling with $\pi N$, $\eta N$, and $\pi\pi N$ reaction 
channels.

Let us now turn to other partial waves. In Table~\ref{tab1}, the extracted 
resonance poles positions ($M_R$) are compared with the bare $N^*$ masses 
($M^0_{N^*}$) of the JLMS model and the 3- and 4-star values listed by 
PDG~\cite{pdg}. With the exception of the $P_{33}(1600)$, $P_{13}$ and 
$P_{31}$ cases, all pole positions listed by the PDG are consistent 
with our results. One possible reason for not finding these poles is 
that their imaginary part may be beyond the $-\text{Im}(M_R) \leq 250$ 
MeV region where our analytic continuation method is accurate and is 
covered in our searches. Another possibility is that these resonances, 
if indeed exist, are perhaps due to the mechanisms which are beyond 
the JLMS model, but are particularly sensitive to these partial waves. 
On the other hand, the possibility that these resonances do not exist 
can not be excluded since the $\pi N$ data are not complete and all 
partial wave analyses involve unavoidable theoretical assumptions. 
For the $F_{35}$ partial wave, we have also analyzed the evolution trajectories
and found that the two poles listed in Table~\ref{tab1} 
 correspond to the same bare state at 2162 MeV. 

In summary, we have applied an analytic continuation method~\cite{ssl09}
 to extract nucleon resonances from a dynamical coupled-channels model
 within which the bare $N^*$ states  were determined
 from fitting the   $\pi N$ scattering data up to $W=2$ GeV~\cite{jlms07}.
 Compared with all previous analysis, the new aspect of this work is to
 study the evolution of resonance pole parameters as a
 function of the coupling to continuum meson-baryon channels.
 Our most important finding is that the two lowest $P_{11}$ nucleon 
 resonances, the Roper $N^*$(1440) and $N^*(1710)$, originate
 from a single bare state. Our finding has an important implication 
in  understanding how nucleon resonances arise in QCD. 
It implies that in some limits in which the coupling to the continuum is
not fully implemented, for example large $N_c$ QCD or quenched lattice QCD,
there could be fewer nucleon resonances. Another possible implication is 
that the bare $N^*$ states, not the resonance poles, determined within 
our model could correspond to hadron structure calculations which 
exclude the coupling with meson-baryon continuum. Further investigations 
of these possibilities as well as related theoretical questions are 
needed to  open a new direction towards understanding nucleon resonances 
and their connection to QCD.
Finally, we mention that our results have confirmed most of the
3- and 4-stars nucleon resonance poles listed by PDG but found
no evidence of two four star resonances, $P_{13}(1720)$,
$P_{31}(1910)$, and one three star one, $P_{33}(1600)$.

\begin{acknowledgments}
 This work is supported by the Japan Society for the Promotion of Science,
Grant-in-Aid for Scientific Research(C) 20540270, 
by the U.S. Department of Energy, Office of Nuclear Physics Division, under
contract No. DE-AC02-06CH11357, and Contract No. DE-AC05-060R23177
under which Jefferson Science Associates operates Jefferson Lab, and
by a CPAN CSD 2007-0042 contract, by Grants No. FIS2008-1661 (Spain).

\end{acknowledgments}

\end{document}